\documentclass[conference]{IEEEtran}

\usepackage{graphicx,mathptmx,amsmath,amsfonts,amsthm,color,hyperref,comment}
\usepackage{enumitem}
\usepackage{enumerate}
\usepackage{algorithmic}

\newtheorem*{example*}{Example}

\def\ben{\begin{eqnarray*}}
\def\een{\end{eqnarray*}}

\def  \GCRC{G_\text{CRC}}
\def  \GPOLAR{G_\text{Polar}}
\def  \MINTER{M_\text{Interleave}}
\def  \H5G{H_\text{CA-Polar}}
\def  \SNR{\text{SNR}}
\def  \BLER{\text{BLER}}
\def  \AB{\text{AB}}
\def  \NQ{Q}
\def  \THRESH{T}
\def  \MYEQ{\stackrel{\text{?}}{=}}

\def  \MERR{ \text{MERR} }
\def  \NORMAL{ \mathcal{N} }
\def  \Ninv{ F_{\NORMAL}^{-1} }

\begin{document}

%---------------------------------------------------------------------------------------------------------------------------

\title{5G NR CA-Polar Maximum Likelihood Decoding by GRAND}

% Guessing random additive noise decoding

\author{
\IEEEauthorblockN{Ken R. Duffy\IEEEauthorrefmark{1}, Amit Solomon\IEEEauthorrefmark{2}, Kishori M. Konwar\IEEEauthorrefmark{2} and Muriel M\'edard\IEEEauthorrefmark{2}\\}
\IEEEauthorblockA{\IEEEauthorrefmark{1}Hamilton Institute, Maynooth University, Ireland.\\}
\IEEEauthorblockA{\IEEEauthorrefmark{2}Research Laboratory of Electronics, Massachusetts Institute of Technology, 
U.S.A.}
}

\maketitle

%------------------------------------------------------------------------------------------------------------------------------------

\begin{abstract}
CA-Polar codes have been selected for all control channel communications in 5G NR, but accurate, computationally feasible decoders are still subject to development. Here we report the performance of a recently proposed class of optimally precise Maximum Likelihood (ML) decoders, GRAND, that can be used with any block-code. As published theoretical results indicate that GRAND is computationally efficient for short-length, high-rate codes and 5G CA-Polar codes are in that class, here we consider GRAND's utility for decoding them. Simulation results indicate that decoding of 5G CA-Polar codes by GRAND, and a simple soft detection variant, is a practical possibility.
\end{abstract}


%------------------------------------------------------------------------------------------------------------------------------------
\vspace{-0.12in}
\section{Introduction}

Polar codes, which were introduced by E. Arikan in 2008, were the
first explicit code construction to be provably channel capacity
achieving ~\cite{Arikan2008,Arikan09}. Their promise of high rates
at short block-lengths led them to be considered for the protection
of 5G control channel communications. Initial results, however,
provided disappointing block error rate (BLER) performance
\cite{Pfister2014,tal2015list}. Subsequent work established that
performance could be substantially enhanced by an additional layer
of redundancy, where data is first coded with a Cyclic Redundancy
Check (CRC) and then Polar coded, leading to CRC Assisted Polar
(CA-Polar) codes.  The standard decoding design is to first list-decode
the Polar code, creating a collection of candidate code-words, and
then select a decoded element by evaluating the CRC of each list
member. This work has resulted in decoders that have shown significant
improvements in BLER, particularly when availing of soft detection
information ~\cite{niu2012crc,tal2015list,balatsoukas2015llr},
resulting in the 5G NR standard adopting CA-Polar codes for all
control communications.

The further development of accurate and computationally efficient
CA-Polar decoders, particularly those that can work in the absence
of soft information, is desirable. Since Shannon's earliest
work~\cite{Shannon48} it has been known that Maximum Likelihood
(ML) decoding is optimal in terms of accuracy for any code construction.
In ML decoding, the decoded word is determined to be the most likely
member of the code-book given the channel output. ML decoding has
not been implemented in practice as it is not computationally
feasible under existing proposals. Instead decoding algorithms are
typically co-designed with particular code-book structures in mind
and heuristically aim to approximately identify an ML decoding
candidate.

An exception to this is the Guessing Random Additive Noise Decoding
(GRAND) framework, which was introduced in 2018~\cite{Duffy18} and
works with any block-code. GRAND identifies an ML decoding, while
GRANDAB (GRAND with Abandonment) \cite{Duffy19}, a variant with 
reduced computational complexity, either identifies an ML decoding
or reports a decoding failure. Both have been theoretically proven
to be capacity-achieving when used with random code-books \cite{Duffy19}.
In contrast to code-book oriented algorithms, GRAND and GRANDAB are
noise-centric, and aim to infer the noise that has occurred on the
channel from which the ML decoding can be deduced. They can decode
CA-Polar codes in a single step, without the need for two separate
docoders.

Unlike most of the more effective existing CA-Polar decoders, GRAND
and GRANDAB are hard detection decoders that solely take demodulated
symbols as input, potentially making them useful for hardware where
soft detection information cannot be provided by the receiver to
the decoder. If symbol reliability information, a simple form of
soft information in which a symbol is marked to be either reliable
or unreliable via, e.g., instantaneous SINR, is available, however,
variants called SGRAND and SGRANDAB that avail of it to improve
decoding accuracy and reduce computational complexity have recently
been introduced \cite{Duffy19a,Duffy19b}. Mathematical results for
random code-books suggest that GRAND and its variants are particularly
appropriate for short, high-rate codes \cite{Duffy19, Duffy19b}.
As in common target operating regimes, raw BER are sufficiently low
to lend themselves to CA-Polar codes that are short-length and
high-rates, the GRAND family of decoders seem to be promising
candidates for 5G NR. Here we report simulation results on 5G
CA-Polar decoding by the GRAND and SGRAND approaches. These results
suggest that accurate decoding 5G communications channel packets
is practically feasible with GRANDAB and SGRANDAB.

\section{Guessing Random Additive Noise Decoding}

The basis for GRAND is that for discrete channels subject to
additive noise the following two algorithms produce the same output,
but that the second one can be computationally feasible for 
high-rate codes while the first one is not. 1) \underline{ML decoding
by brute force:} given a received block of $n$ demodulated symbols,
$y^n$, and a channel noise model, compute the likelihood that each
code-word in the code-book, $c^{i,n}$, was transmitted given $y^n$
was received, and define the decoded element, $c^{*,n}$, to be
one with the highest likelihood. 2) \underline{ML decoding by GRAND:} taking
noise sequences, $z^n$, in order from most likely to least
likely based on the channel model, subtract them (in the Galois field
of the symbols) from the received sequence $y^n$ and query if what
remains, $y^n\ominus z^n$, is a member of the code-book, reporting
the first instance, $c^{*,n}$, where that is true. 

To GRAND, GRANDAB adds a counter for the number of noise guesses,
i.e. code-book membership queries, made, and abandons guessing if
more than a set number has been exceeded. It has been proven that
GRANDAB is capacity achieving when used with random code-books so
long as the abandonment threshold is set correctly \cite{Duffy19}.

For a linear code construction with parity check matrix $H$, one
can test if a string, $y^n$, is a member of the code-book by a single
matrix multiplication and comparison, $H(y^n)^T\MYEQ(0^n)^T$. All 5G
NR codes are linear and binary, so that the appropriate field
is $\text{GF}_2$ (i.e. mod 2). Control channel communications are
first coded with a CRC, then interleaved if they are downlink
communications, and finally Polar coded prior to modulation
and transmission. All of those operations are linear, so
that an input information word consisting of $k$ bits, $x^k$, is
transformed into an $n$-bit code-word, $c^n$, by the linear map
\begin{align*}
c^n = x^k\GCRC\, \MINTER \,\GPOLAR,
\end{align*}
where $\GCRC$ is the generator matrix for the CRC, $\MINTER$ is the
interleaving matrix (the identity for uplink communications), and
$\GPOLAR$ is the generator for the Polar code.

As most decoders are code-book centric, two distinct decoders would
normally need to be employed: one for the Polar Code and one for
the CRC. As the GRAND algorithms are code-book agnostic, they can
treat the CA-Polar code as a single linear code and decode both
simultaneously by using the corresponding parity check matrix $\H5G$,
and testing code-book membership, $\H5G(y^n)^T\MYEQ(0^n)^T$. As a
result, the decoder provides a code-word that is consistent with
both the CRC and Polar code without the need for two separate
decoders. Pseudo-code for GRANDAB can be found in
Fig.~\ref{alg:pseudo-code}.

\begin{figure}
\hrule
\noindent
\begin{algorithmic}
\STATE {\bf Inputs}: $y^n$, $H$, $\THRESH$
\STATE {\bf Output}: $c^{*,n}$, $d$, $\NQ$
\STATE $d\leftarrow 0$, $\NQ\leftarrow 0$.
\WHILE{$\NQ\leq\THRESH$}
\STATE $z^n\leftarrow$ next most likely noise sequence
\STATE $\NQ\leftarrow\NQ+1$ 
 \IF{$H(y^n\ominus z^n)^T = (0^n)^T$}
\STATE  $c^{*,n}\leftarrow y^n\ominus z^n$
\STATE  $d\leftarrow1$
\STATE{\bf return} $c^{*,n}$, $d$, $\NQ$
\ENDIF
\ENDWHILE
\STATE {\bf return } $\perp$, $d$, $\NQ$ /* failed to decode due to abandonment */
\STATE
\STATE
\hrule
\end{algorithmic}
\caption{Sketch of GRAND and GRANDAB. Given a
channel output $y^n$, a block code's parity check matrix $H$, and
a querying abandonment threshold $\THRESH$ ($=\infty$ for GRAND),
if a code-book element $c^{*,n}$ is identified before a number of
code-book queries, $\NQ$, corresponding to the guesswork threshold is
exceeded, it is reported along with successful decoding, $d=1$;
otherwise an abandonment failure is reported, $d=0$.}
\label{alg:pseudo-code}
\vspace{-0.5cm}
\end{figure}

\section{5G CA-Polar Codes by GRAND}

Core to GRAND's accuracy is querying noise sequences in order from
most likely to least likely. An interleaved transmission corresponds
to a Binary Symmetric Channel (BSC) \cite{Heegard1999} where the
binary channel output $y^n$ can be written as the binary input
code-word, $c^n$, plus independently binary additive channel noise
$z^n$, $y^n=c^n\oplus z^n$. Thus we use the likelihood ordering
determined by the BSC where the most likely sequence is all zeros,
followed by each of those with one bit flip in any order, followed
by those with two bit flips in any order, etc.

In our implementation of the GRANDAB, we abandon guessing if no
element of the code-book has been identified after testing all
sequences with up to $\AB$ bit flips. If the noise introduces more
than $\AB$ bit flips, as no such error sequences will be queried,
this necessarily results in an error either due to abandonment or
an erroneous decoding. Using GRAND, we first establish that an
appropriate value of $\AB$ is solely a function of the code-book
and not channel conditions.

\begin{figure}[h]
\begin{center}
\includegraphics[width=0.38\textwidth]{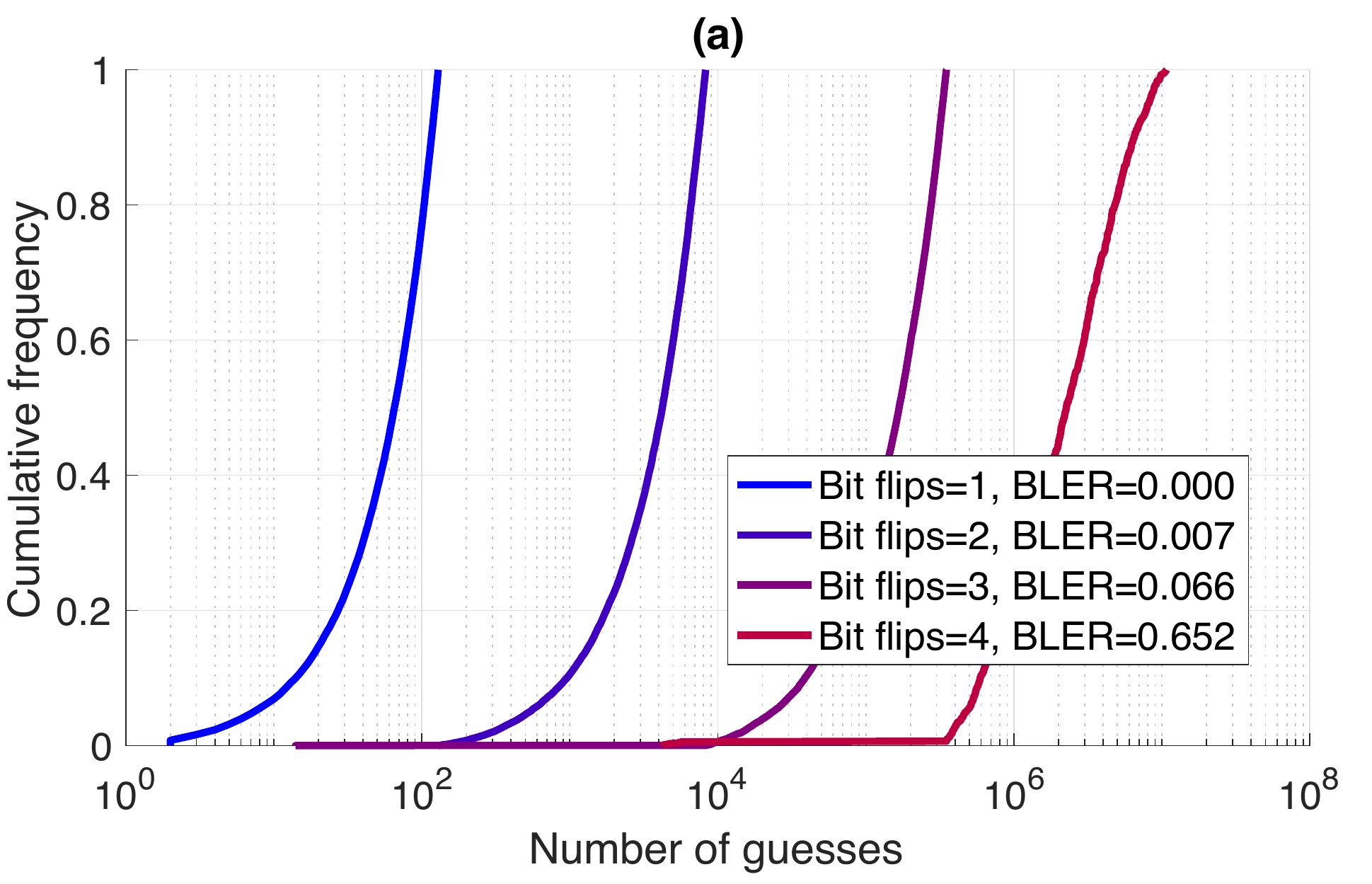}
\includegraphics[width=0.38\textwidth]{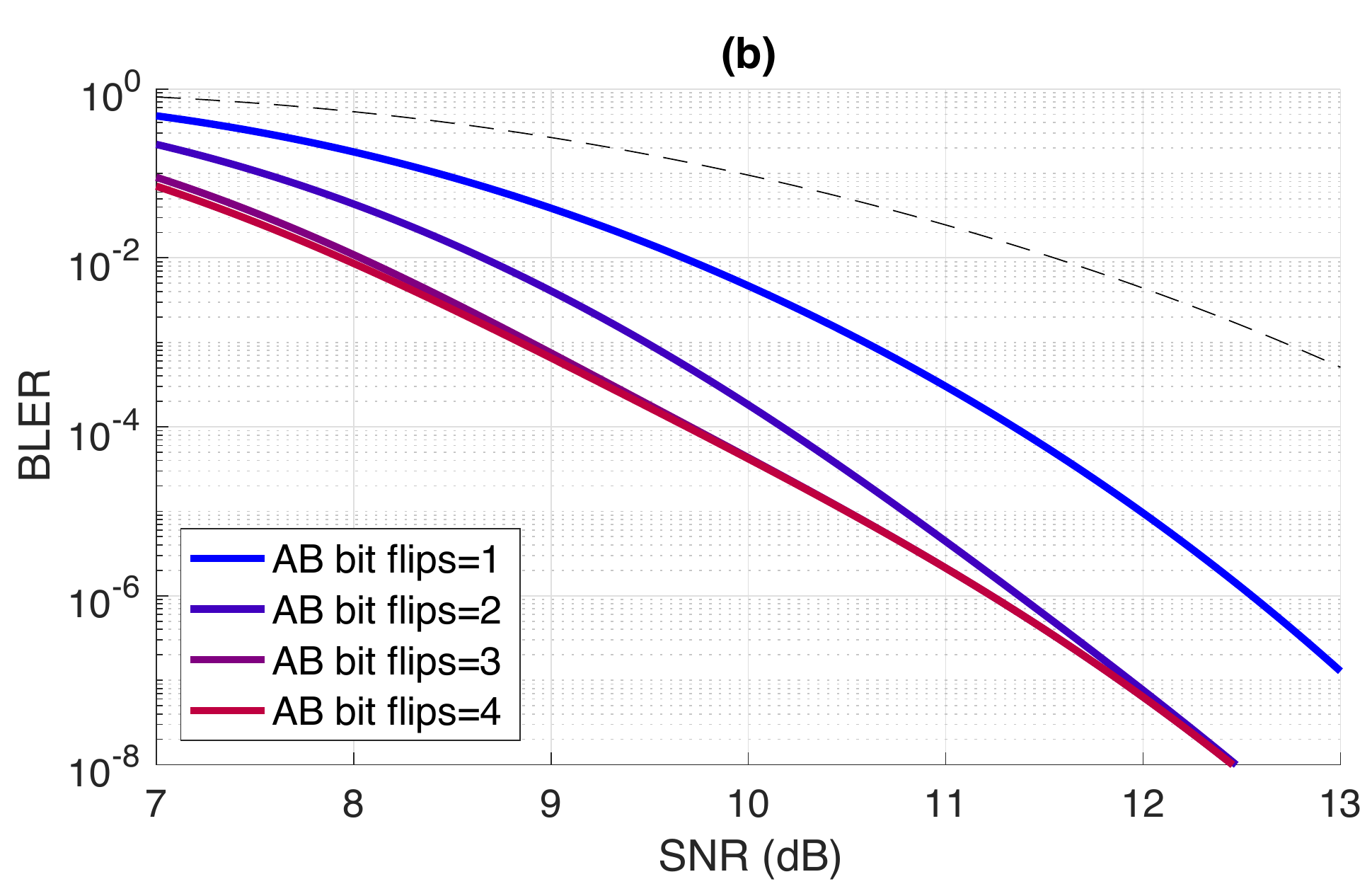}
\includegraphics[width=0.38\textwidth]{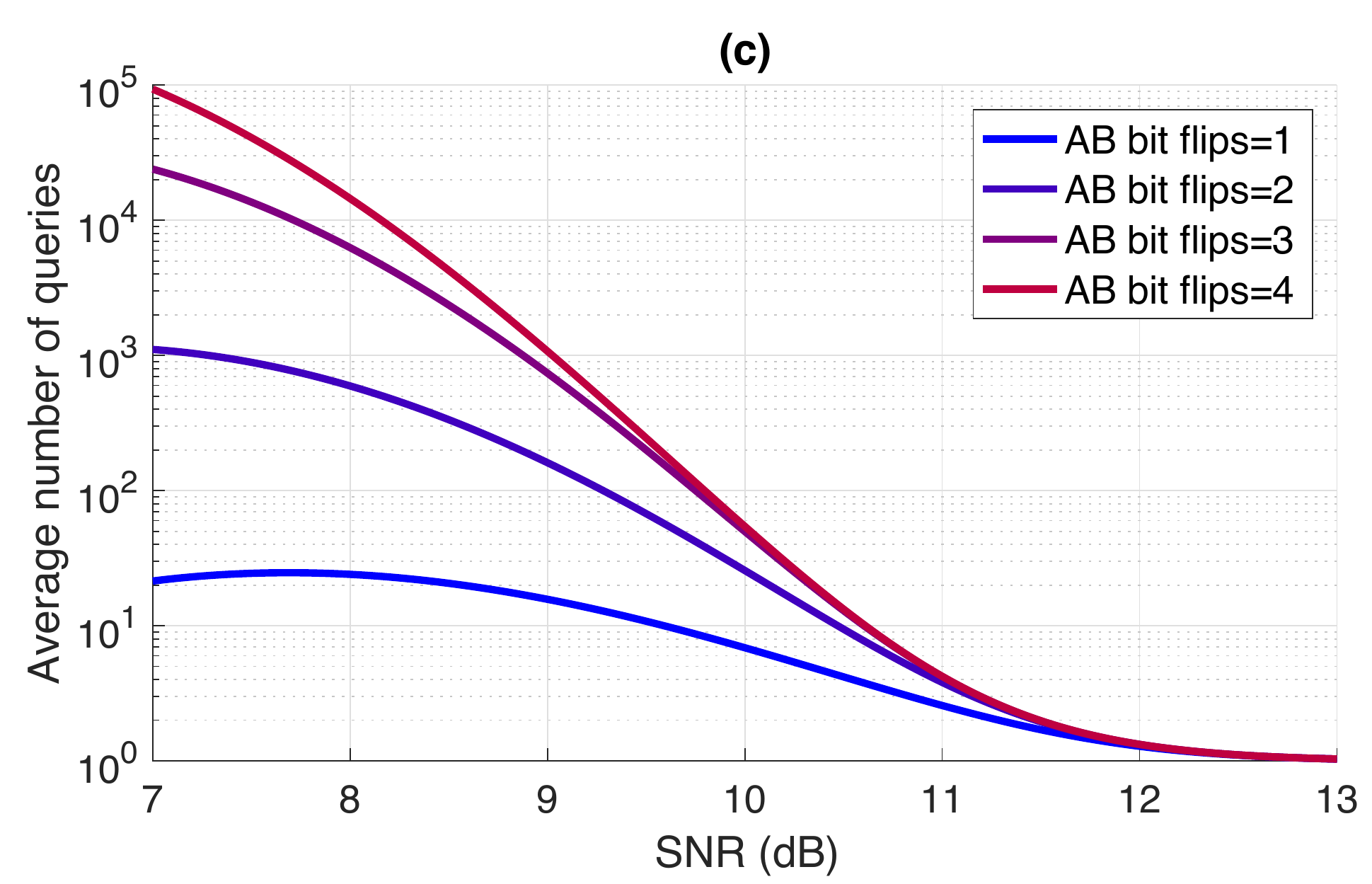}
\end{center}
\caption{5G NR uplink CA-Polar [128,105] decoding by GRAND and
GRANDAB. (a) Conditioned on the number of channel noise bits flips,
the cumulative distribution of how many code-book queries
are made by GRAND before a decoding is identified. Reported in the
legend is the block error rate given that number of bit flips. (b)
GRANDAB BLER vs SNR in an AWGN BPSK channel. The dashed black line
indicates uncoded block error rate (i.e. the likelihood that one or more
bits are flipped in a 128 bit block). The solid lines indicate
BLER for GRANDAB given the abandonment after querying all noise
sequences with weight up to $\AB$. (c) Average number of code-book
queries vs SNR until a decoding is found or GRANDAB
abandons guessing.
}
\label{fig:1}
\end{figure}

An example high rate CA-Polar code in the 5G NR standard is for the
uplink where $k=105$ information bits are converted results into
$n=128$ transmitted bits per block, giving a rate of $0.82$. Figure
\ref{fig:1} (a) plots the cumulative distribution of the number of
code-book queries, $\NQ$, made by GRAND until a decoding is identified
conditioned on the number of bit flips due to noise. For each
number of conditioning bit flips, $b$, these plots were created by
simulation of a large number of noise sequences with $b$ bit flips
randomly located in $1,\ldots,n$. For $b=1,2,3$, these distributions
have similar structure as they primarily correspond to the number
of code-book queries until a correct decoding is identified. When
$b=4$, the distribution looks distinct because many of the decodings
are erroneous.

Also shown in the legend of Fig. \ref{fig:1} (a) is GRAND's BLER
conditioned on that number of bit flips. As GRAND identifies an ML
decoding, a more accurate decoding algorithm is not possible with
only hard detection information. If a single bit is flipped, the
conditional BLER is zero and the decoding is always correct. If
four bits are flipped, however, the ML decoding is in error for
approximately $65\%$ of ML decodings as identified by GRAND,
indicating that, with high probability, the minimum distance of the
code is 3 or 4. In a concatenated code design with an optimial
outer code, the additional redundancy required to correct an erroneous
decoding is just over twice that required to correct an erasure
\cite{Roth2006}.  As a result, if, for a given number of bit flips,
the ML decoding has a conditional BLER $>33\%$ of decodings, the
decoder would be better off not doing the work required to determine
the ML decoding, but instead report an erasure by abandoning via
GRANDAB. By this rationale, for the CA-Polar [128,105] code, GRANDAB
should abandon decoding if no code-book element is identified after
up to all Hamming weight sequences of $\AB=3$ have been queried.

From the data contained in Figure \ref{fig:1}, we can efficiently
determine GRANDAB's BPSK AWGN BLER and complexity, in terms of the number of
code-book queries, via importance sampling, the law of total
probability, and the law of total expectation. That is, with $B$
being the number of bit flips due to noise, we have
\begin{align}
\label{eq:sim}
\BLER &= \sum_{b=0}^\AB P(\text{error}| B=b) P(B=b) + P(B>\AB)\\
E(\NQ) &= \sum_{b=0}^\AB E(\NQ| B=b) P(B=b) + \left(\sum_{b=0}^\AB {n \choose b}\right)P(B>\AB). \nonumber
\end{align}
Assuming binary phase shift keying (BPSK) to $\{-1,+1\}$, and AWGN
with variance $\sigma^2$, the signal-to-noise ratio is $\SNR =
-10\log_{10}\sigma^2$ in dB (note that there are other definitions
of SNR, e.g. Eb/N0, which may complicate direct comparison) and
the probability of a bit flip is $p=P(\sigma\NORMAL>1)$,
where $\NORMAL$ denotes a normal random variable with mean $0$ and
variance $1$.  $B$ is distributed as a binomial random variable
with probability $p$ on $n$ trials, where $n$ is the code-length.
Thus to evaluate the BLER and the computational complexity, we use
the conditional information in Fig \ref{fig:1} (a) to estimate
$P(\text{error}| B=b)$ and $E(\NQ| B=b)$, and then use eq.
\eqref{eq:sim}.

GRANDAB's BLER vs SNR is shown in Fig \ref{fig:1} (b). The dashed
black line indicates the uncoded BLER (i.e. the probability that a
single bit in a $128$ bit block is flipped by noise), while the
colored lines correspond to GRANDAB with a range of abandonment
weights. Consistent with earlier logic, the BLER of GRANDAB with
$\AB=3$ or $4$ is essentially identical, indicating that there is
no gain in BLER to be had with this CA-Polar code by
seeking a decoding where four or more bits have been flipped. In
comparison to abandonment after only one or two bit flips, $\AB=1$
or $\AB=2$, however, both have substantially better BLER. This
reflects the fact that with $\AB=3$, GRANDAB is essentially providing
all of the merits of an ML decoder, but with an a priori upper
bound on decoding complexity.

In terms of complexity, Fig \ref{fig:1} (c) reports the average
number of code-book queries until GRANDAB identifies a decoding.
A standard operational regime would seek a BLER of $<10^{-3}$, which
is achieved with this code for a SNR of $9$dB or higher whereupon
it is over two orders of magnitude less than the uncoded BLER. For
those SNRs, GRANDAB with $\AB=3$ performs fewer than $10^3$ code-book
queries per decoding on average, i.e. $<10$ code-book queries per
decoded bit. As code-book queries can readily be parallelized, this
holds significant promise for both software and hardware implementations.

\begin{figure}[h]
\begin{center}
\includegraphics[width=0.38\textwidth]{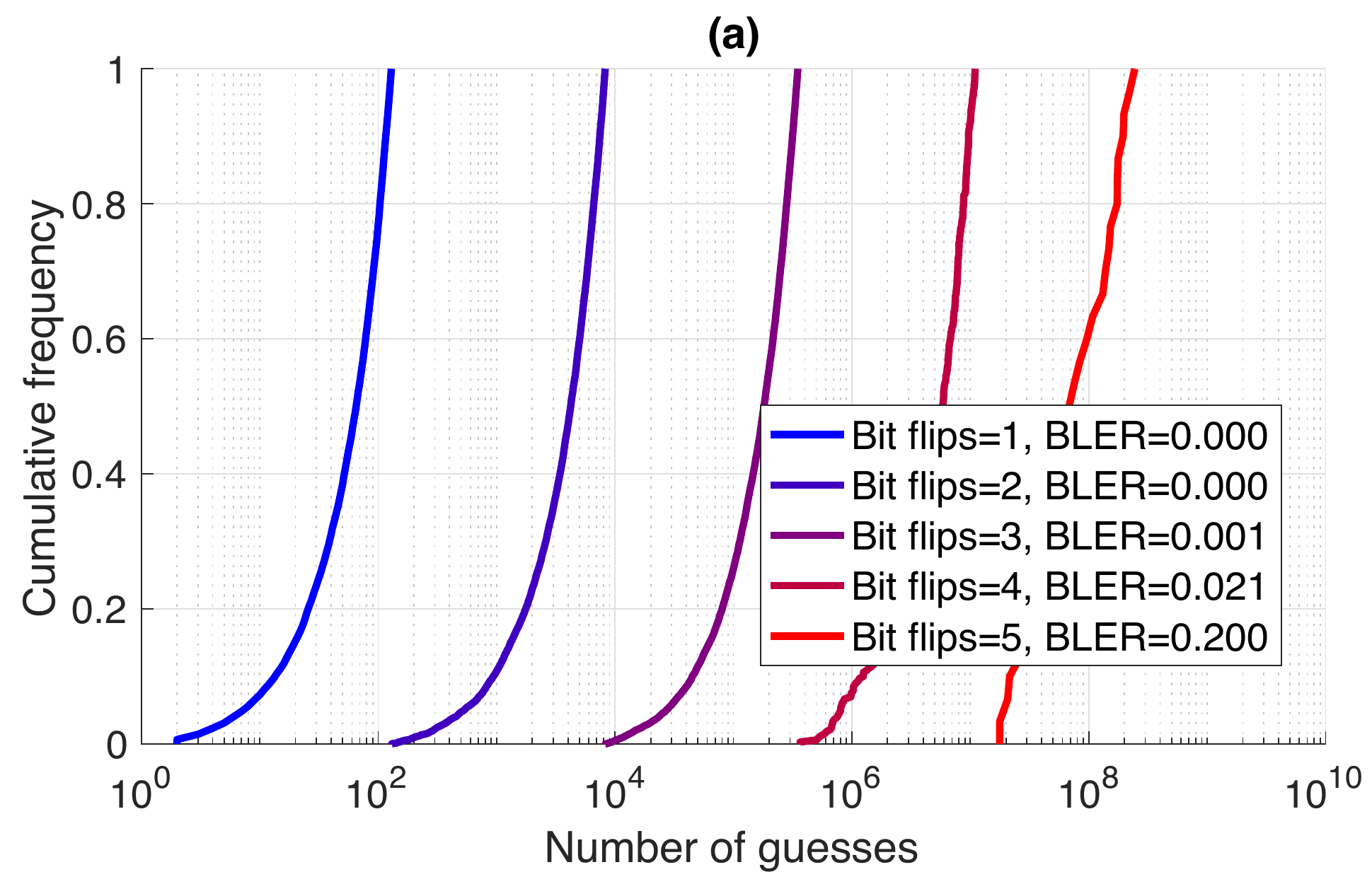}
\includegraphics[width=0.38\textwidth]{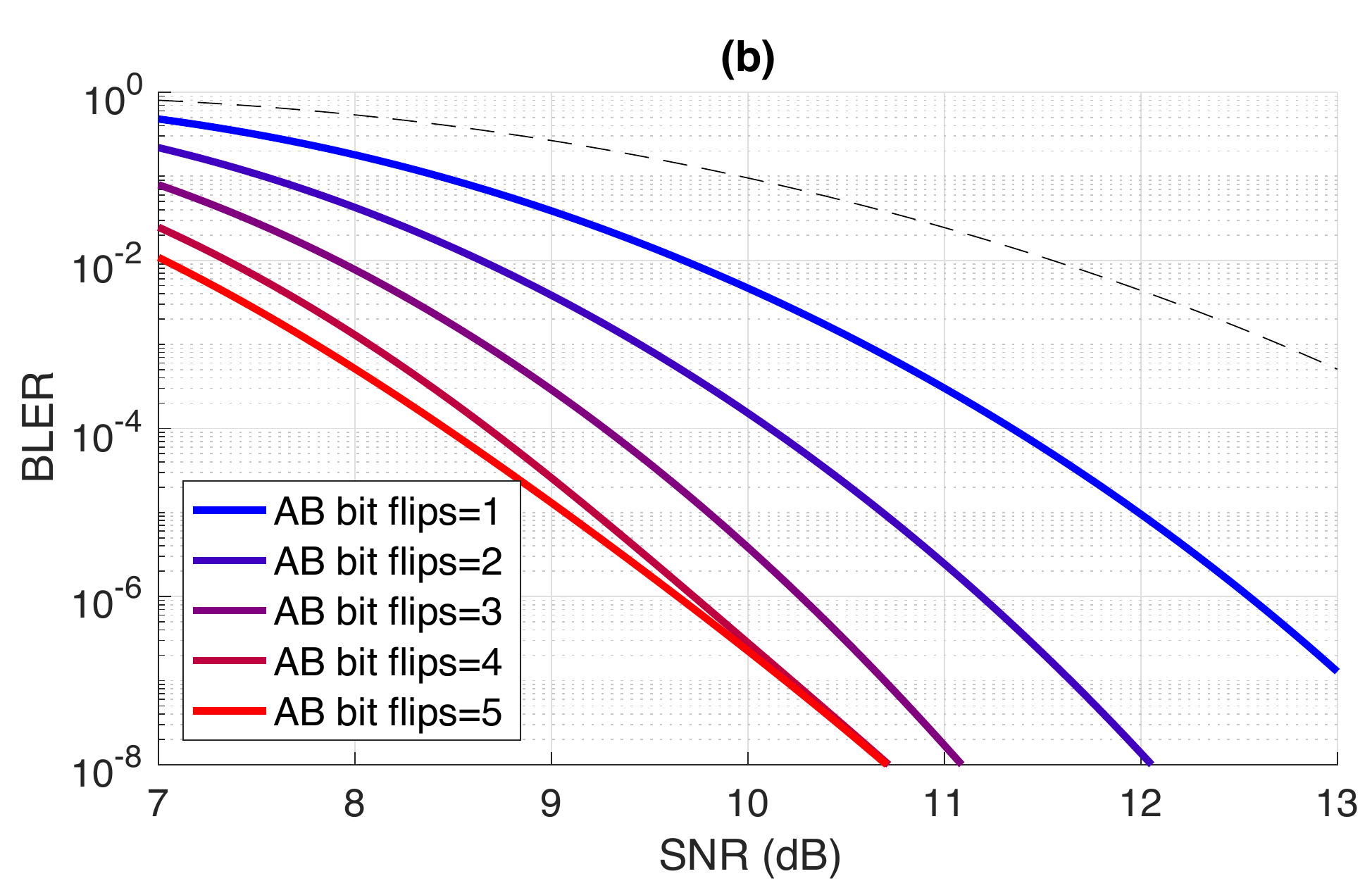}
\includegraphics[width=0.38\textwidth]{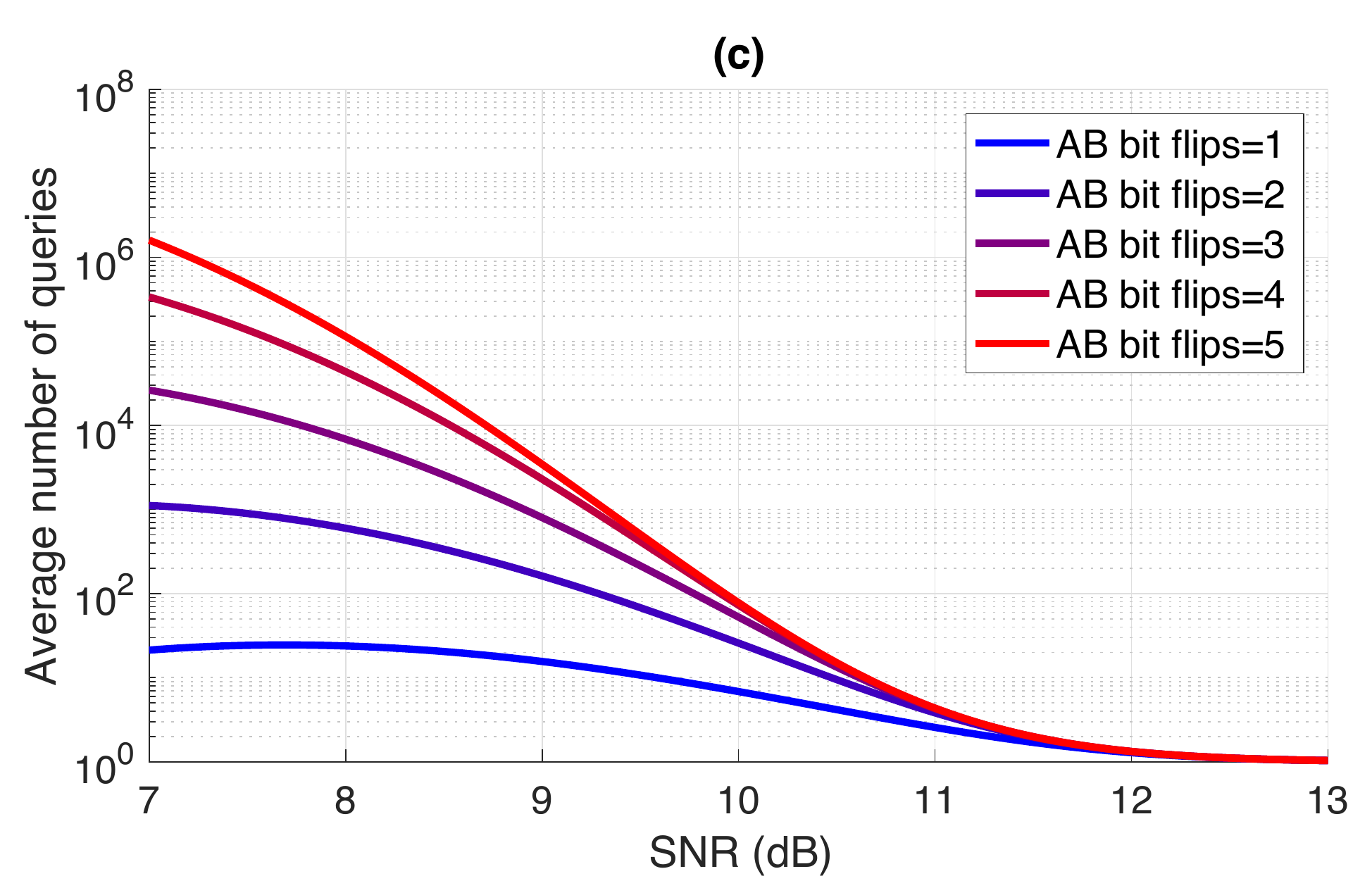}
\end{center}
\caption{5G NR downlink CA-Polar [128,99].
(a-c) As in Fig. \ref{fig:1}.
}
\label{fig:2}
\end{figure}

For comparison, another high rate CA-Polar code encodes $k=99$
information bits into $n=128$ bits for the downlink, giving a rate
of $0.77$.  Fig \ref{fig:2} plots analogous results to those of the
[128,105] uplink code shown in Fig \ref{fig:1}. In order to fully
utilize the redundancy available in this lower-rate code, the
abandonment has to be pushed out to a Hamming weight of $\AB=4$,
at a cost of extra work for the decoder. This potentially
counter-intuitive result reflects that the GRAND approach favors
high rate codes, which is a consequence of the fact that the
complexity of GRAND algorithms can only decrease as code-rates
increase.

These results suggest that GRANDAB holds promise as an accurate and
efficient hard-detection CA-Polar decoder. If in addition quantized
symbol reliability information can be provided by the receiver to
the decoder, recent theoretical results suggest that further
improvements in BLER and decreases in complexity are possible
\cite{Duffy19a,Duffy19b}, which we now explore.

\section{5G NR CA-Polar Codes by SGRAND}

\begin{figure}[h]
\begin{center}
\includegraphics[width=0.38\textwidth]{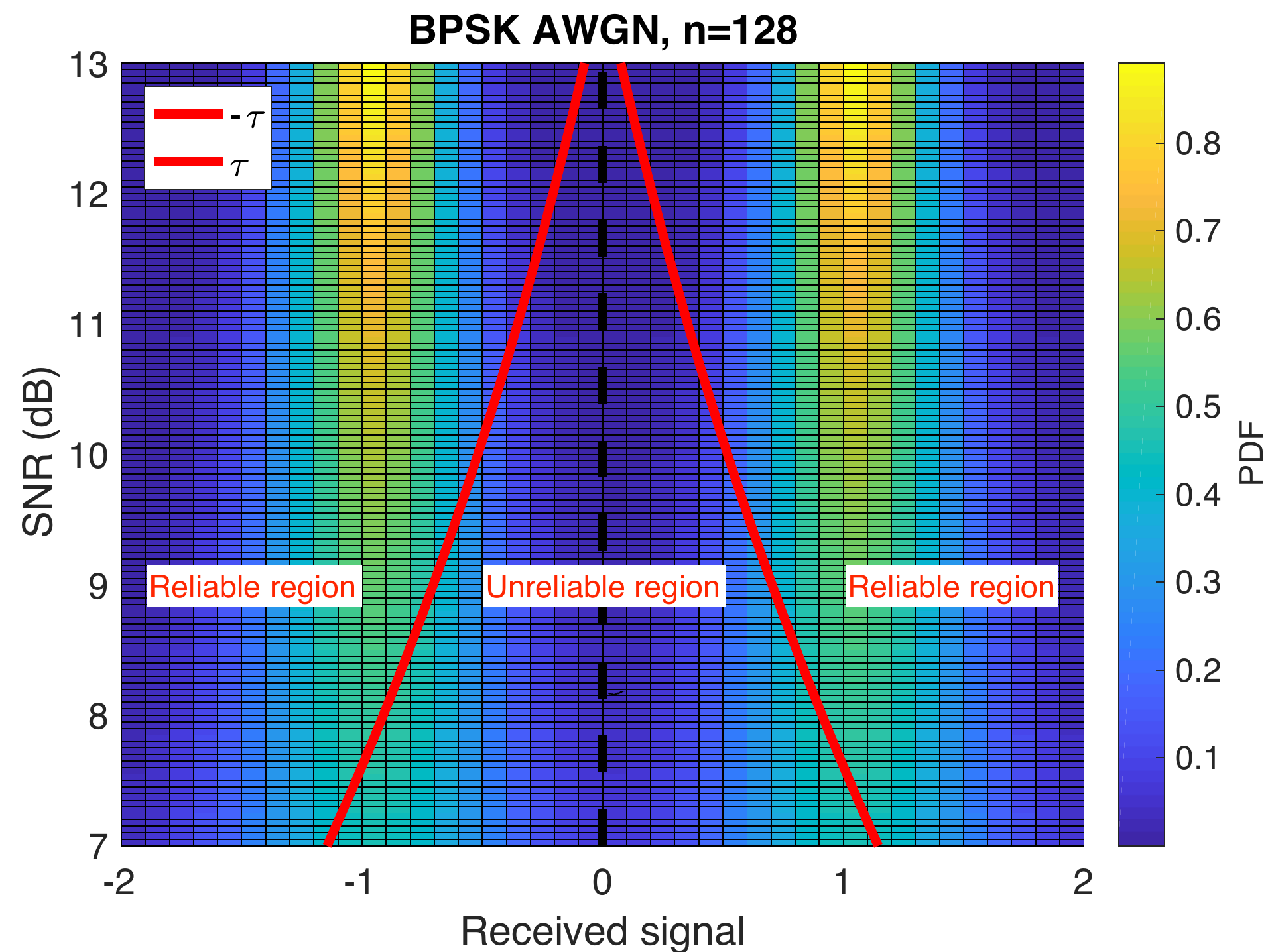}
\end{center}
\caption{BPSK AWGN symbol reliability. Bits are modulated to
$\{-1,+1\}$ and are subject to AWGN. Shown is a heat map
of the probability density of the received signal (x-axis) as a
function of SNR (y-axis). The code-book independent quantization
of soft detection information in SGRAND sees demodulated bits
corresponding to signals received outside of the red region
marked as reliable, while those that within are marked as unreliable.
The region is chosen so that the probability that any bit in a
transmitted $n=128$ block is marked as reliable in error (e.g. the
received signal is left of the red-line, while a $+1$ was transmitted)
has probability $\MERR=10^{-4}$.
}
\label{fig:3}
\end{figure}

In BPSK, each coded bit is communicated via the phase of a continuous
wave that is impacted by noise before reception and demodulation.
Fig. \ref{fig:3}) provides a heat-map plot
of the probability density of the received signal as a function of
the SNR in an AWGN model. In the hard detection setting of the
previous section, the received signal is demodulated to a $0$ or
$1$ depending on whether it is to the left or right of the vertical
dashed black line, and the resulting bits passed to the decoder.
Hard detection decoders execute solely based on that information.

Soft detection decoders attempt make further use of the received
signal to better inform their decoding, e.g.
\cite{Coo88,HLY02,KNH97,GS99,XCB17}. Incorporating soft detection
information results in improved accuracy, but typically at the cost
of increased computational complexity in the decoding process.
Moreover, provision of soft detection information, such as instantaneous
SINR, by a receiver to a decoder requires the passing of real-valued
data, which is costly in terms of memory and I/O. An alternative
approach, first considered within Chase decoding \cite{Cha72,MH17},
is for the receiver to quantize the received signal and provide
symbol reliability information where, in addition to the hard-demodulated
symbols, $y^n$, the receiver passes a binary string, $s^n$, that
labels symbols as reliable or unreliable. Requiring only one
additional bit per received symbols, symbol reliability information
is much less costly to provide to the decoder.

SGRAND and SGRANDAB \cite{Duffy19a,Duffy19b} expect such symbol
reliability information. For BPSK, we created that information by
setting a threshold for the received signal, $\pm\tau$, above or
below which one is confident that the transmitted phase was $\pm
1$. We define the Mask Error Rate, $\MERR$, to be error rate due
to erroneously marking one or more bits in a block as reliable when
they are not. The threshold, $\tau$, is determined as a function
of the SNR, the code-length, and $\MERR$. As the probability a bit
is erroneously labeled as reliable when it is incorrect is
$P(\sigma\NORMAL>1+\tau)$, we have $\MERR = 1-P(\sigma\NORMAL\leq
1+\tau)^n$.  Hence we set $\tau = \sigma \Ninv((1-\MERR)^{1/n})-1$,
where $\Ninv$ is the inverse of a Normal distribution. For $n=128$
and $\MERR=10^{-4}$, the red lines in Fig. \ref{fig:3} illustrate
the masked region as a function of SNR.

\begin{figure}
\hrule
\noindent
\begin{algorithmic}
\STATE {\bf Inputs}: $y^n$, $H$, $\THRESH$, $s^n$,
\STATE {\bf Output}: $c^{*,n}$, $d$, $\NQ$
\STATE $d\leftarrow 0$, $\NQ\leftarrow 0$.
\WHILE{$\NQ\leq\THRESH$}
\STATE $z^n\leftarrow$ zeros where $s^n$ is zero and the next most likely noise sequence
mapped to the ones of $s^n$ 
\STATE $\NQ\leftarrow\NQ+1$ 
 \IF{$H(y^n\ominus z^n)^T = (0^n)^T$}
\STATE  $c^{*,n}\leftarrow y^n\ominus z^n$
\STATE  $d\leftarrow1$
\STATE{\bf return} $c^{*,n}$, $d$, $\NQ$
\ENDIF
\ENDWHILE
\STATE {\bf return } $\perp$, $d$, $\NQ$ /* failed to decode due to abandonment */
\STATE
\STATE
\hrule
\end{algorithmic}
\caption{Sketch of SGRAND and SGRANDAB. Given the
same inputs as in Fig. \ref{alg:pseudo-code} and $s^n$, a binary
mask of length $n$ with $l^n$ ones in locations indicating unreliable
symbols, if a code-book element $c^{*,n}$ is identified before a
number of code-book queries corresponding to the guesswork threshold
is exceeded, it is reported along with successful decoding, $d=1$;
otherwise an abandonment failure is reported, $d=0$.}
\label{alg:pseudo-code2}
\vspace{-0.5cm}
\end{figure}

Armed with this symbol reliability information, pseudo code for
SGRANDAB is given in Fig.~\ref{alg:pseudo-code2}. SGRANDAB proceeds
as GRANDAB, but with subtraction of noise sequences restricted to
only bits marked as unreliable. The result is a more targeted
querying on an effectively smaller code, leading to better decoding
accuracy and less complexity. As with the hard detection decoders,
BLER and complexity performance can be determined by the importance
sampling approach, first conditioning on mask-lengths (i.e. number
of unreliable bits) and then on the number of bit flips. In the
BPSK AWGN setting, the probability that a bit is marked as unreliable
is $q=P(\sigma\NORMAL\leq -1+\tau)-P(\sigma\NORMAL\leq -1-\tau)$,
while the conditional probability that a bit is flipped given it
is marked as unreliable is $p=\left(P(\sigma\NORMAL\leq
1+\tau))-P(\sigma\NORMAL\leq1)\right)/q$.

\begin{figure}[h]
\begin{center}
\includegraphics[width=0.38\textwidth]{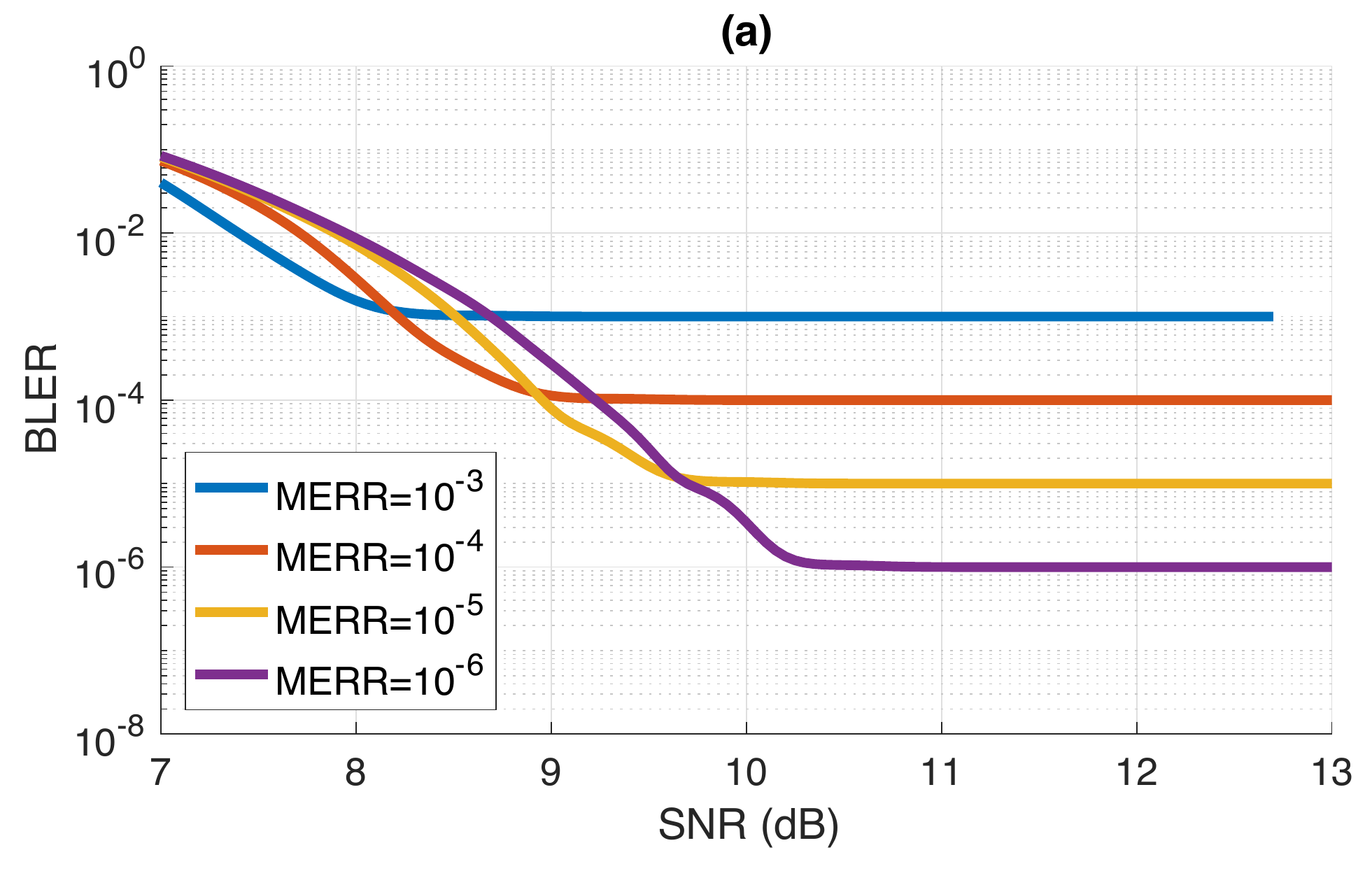}
\includegraphics[width=0.38\textwidth]{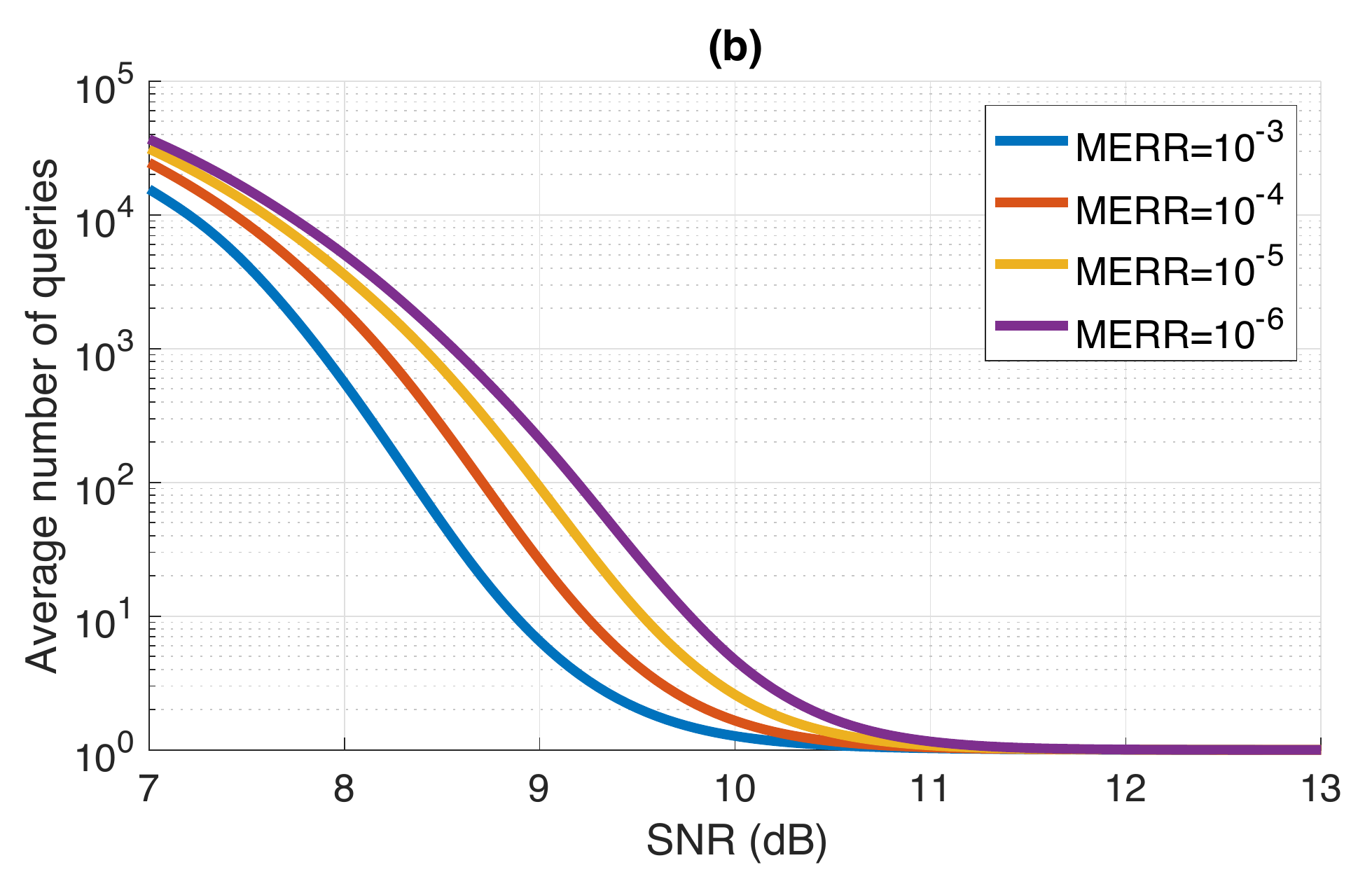}
\end{center}
\caption{5G NR CA-Polar $[128,105]$ decoding by SGRANDAB.  For
different values of target mask error rate, $\MERR$, accuracy and
complexity, where the guesswork threshold is set so as to allow
all three bit flip sequences be queried for a mask of 128 bits.
(a) BLER vs SNR. (b) Average number of code-book queries until
a decoding is identified or SGRANDAB abandons guessing. 
}
\label{fig:4}
\end{figure}

For the $[128,105]$ CA-Polar code considered previously, Fig.~\ref{fig:4}
reports the performance of SGRANDAB where the guesswork threshold
is set to allow the querying of all three bit flip noise sequences
for a mask of 128 bits. For a range of values of $\MERR$, the target
mask-error probability, Fig.~\ref{fig:4} (a) reports BLER vs SNR.
With a $\MERR=10^{-4}$, a BLER of $10^{-4}$ is achieved from a SNR
of 9 for SGRANDAB, in comparison to a SNR of 9.75 for GRANDAB as
provided in Fig. \ref{fig:2} (b). As well as a gain in decoding
accuracy, SGRANDAB provides a significant reduction in decoding
complexity due to the targeted nature of its querying where at an
SNR of 9 for the target $\MERR=10^{-4}$, this BLER is achieved with
fewer than $12$ code-book queries per decoding on average, in
comparison to $110$ for the hard detection equivalent. 

\section{Conclusions}

Due to their high rates at short code-lengths, CA-Polar codes have
been selected for all control channel communications in 5G NR.
Current designs see list decoding, typically with soft detection
information, of the Polar code followed by use of the CRC to select
a decoding. Here we have established that the hard detection
algorithms GRAND and GRANDAB, which identify optimal ML decodings
and decode both codes in a single step, offer a viable alternative
method. If a receiver can pass symbol reliability information to
the decoder, use of SGRANDAB can increase decoding precision with
reduced computational complexity. Taken together, the GRAND approach
seems promising for use with high-rate 5G NR CA-Polar codes.  Further
work may consider ancillary issues such as rate matching and recovery,
or further use of more detailed soft detection information.

\bibliographystyle{IEEEtran}
\bibliography{SGRAND}

\end{document}